\documentclass{article}
\pdfoutput=1 
\usepackage[T1]{fontenc}
\usepackage[utf8]{inputenc}
\usepackage{ismir}
\usepackage{amsmath,cite,url}
\usepackage{xcolor}
\usepackage{graphicx}
\usepackage{booktabs}
\usepackage{color}
\usepackage{comment}
\usepackage{tikz}
\usetikzlibrary{positioning, arrows.meta}
\usepackage{pifont}    

\title{The Internet Archive Music Dataset}

\multauthor
  {Paraskevas Stamatiadis$^1$ \hspace{1cm} Bernardo V. Miranda$^1$ \hspace{1cm} Clémentine Berger$^1$}
  {{\bf Gaël Richard$^1$ \hspace{1cm} Mathieu Fontaine$^1$ \hspace{1cm} Slim Essid$^\dag$}\\
  $^1$ LTCI, Télécom Paris, Institut Polytechnique de Paris, France\\
  {\tt\small stamatiadis@telecom-paris.fr} \hspace{1cm}
  {\tt\small bernardo.vieira@telecom-paris.fr} \\
  {\tt\small clementine.berger@telecom-paris.fr}}

\def\authorname{P. Stamatiadis, B. V. Miranda, C. Berger, G. Richard, M. Fontaine, S. Essid}

\begin{document}

\maketitle

\begin{abstract}
We introduce the Internet Archive Music Dataset (IAMD), a large-scale collection of captioned music segments derived from the Internet Archive. To the best of our knowledge, IAMD constitutes the largest publicly available music–caption dataset to date with over 34,000 hours of audio, providing a valuable benchmark for training and evaluating music understanding and generative models. The dataset is built from content declared to be distributed under Creative Commons licenses, and cross-referencing with MusicBrainz is done to improve license information reliability. To annotate IAMD, we present an automatic captioning pipeline that augments base captions produced by an audio-language model (ALM) with textual metadata sourced from the Internet Archive and imputed metadata obtained using audio classification models. Caption quality is assessed objectively and subjectively, and results indicate that the annotation pipeline is reliable and does not degrade caption quality with scaling.
\end{abstract}

\section{Introduction}\label{sec:introduction}
\begingroup
\renewcommand{\thefootnote}{\fnsymbol{footnote}}
\footnotetext[2]{This work is independent from S. Essid's role at NVIDIA.}
\endgroup
In recent years, both music understanding models and generative models for music have seen substantial performance gains~\cite{evans2025stable, evans2024long, goel2025audioflamingo, ghosh2026music, tal2024joint}. Much of this progress is driven by scaling; as model capacity and computational resources increase, performance on key tasks tends to improve, as reflected in recent deep learning trends~\cite{peebles2023scalable}. However, scaling model size requires increasing dataset size accordingly. This holds for large audio-language models (LALMs) and text-to-audio models in the broader audio domain, but also for music understanding and text-to-music models in the music domain.

Such models are typically trained on pairs of audio and textual descriptions, and scaling datasets of this kind for music is particularly challenging. First, there is the difficulty of captioning large amounts of data, which is also present for general audio and usually requires setting an automatic annotation pipeline due to the time and cost constraints of human labeling. Second, music data should be gathered without copyright infringement, meaning that publicly available datasets should rely on openly licensed audio.

Due to these constraints, it is common to train large multimodal models for music on private datasets sourced from music licensing platforms such as Pond5~\cite{tal2024joint, copet2023simple} and AudioSparx~\cite{evans2024long}. The lack of large-scale music-caption datasets built on openly licensed data remains a significant limitation that hinders reproducible research for music understanding and music generative models. Large-scale, openly licensed music-caption datasets are crucial for reproducible research in these tasks, as their absence makes it difficult to disentangle model improvements from dataset specific effects. 

In this context, we introduce the Internet Archive Music Dataset (IAMD), a large-scale collection of captioned music segments derived from the Internet Archive (IA)~\cite{archive2014about}, a nonprofit digital library containing millions of books, videos, and audio recordings. With around 34.8k hours of captioned music, divided into 4.1M audio segments extracted from 548k files, IAMD is, to the best of our knowledge, the largest music-caption dataset to date (see Table~\ref{tab:datasets_size} for a comparison with existing datasets). The dataset is constructed from IA uploads declared by users to be distributed under Creative Commons (CC) licenses, resulting in a large-scale resource with an emphasis on open licensing. To further improve license reliability, a subset of the data is cross-referenced with the MusicBrainz\footnote{\url{https://musicbrainz.org/}} database, allowing potentially copyrighted material to be flagged through external metadata signals.
\begin{table}[t]
    \small
    \centering
    \setlength{\tabcolsep}{5pt}
    \begin{tabular}{lcc}
        \toprule
        \textbf{Dataset} & \textbf{\# Captions} & \textbf{\# Hours} \\
        \midrule
        MusicCaps~\cite{agostinelli2023musiclm} & 5,521 & 14\\
        Jamendo FMA Captions~\cite{lanzendorfer2025coarse} & 122,480 & 1K\\
        LP-MusicCaps~\cite{doh2023lp} & 513,977 & 4.5K\\
        JamendoMaxCaps~\cite{roy2025jamendo} & 1.8M & 14.5K \\
        \textbf{IAMD} & 4.1M & 34K \\
        \bottomrule
    \end{tabular}
    \caption{Comparison of music-caption datasets by scale. Hours are coarsely rounded for readability.}
    \vspace{-0.3cm}
    \label{tab:datasets_size}
\end{table}

Leveraging a two-stage automatic captioning pipeline, IAMD achieves caption quality comparable to existing large-scale datasets and not far behind that of human-annotated benchmarks, as demonstrated by objective metrics. This suggests that IAMD scales without compromising caption quality, providing a robust resource for training and evaluating both music understanding and generative models. The code\footnote{\url{https://github.com/bvm810/IAMD}} for data collection, filtering, license verification, and automatic captioning is released along with the dataset, supporting transparency, reproducibility, and future extensions. Download links, caption examples, dataset statistics, and further details on our captioning pipeline can be found in our companion website\footnote{\url{https://adasp.telecom-paris.fr/s/iamd}}.
\section{Related Work}\label{sec:related-work}
Music Information Retrieval (MIR) encompasses a broad range of music analysis tasks, including classification~\cite{nam2018deep}, recommendation~\cite{deldjoo2024content}, music understanding~\cite{goel2025audioflamingo,li2026tiny,salganik2026musicsem}, and generative modeling~\cite{evans2024long,liu2024audioldm,gong2025ace}. In recent years, the rapid advancement of state-of-the-art (SOTA) deep learning models has made access to well-annotated data at scale a critical requirement. Historically, however, the community has struggled with limited access to such resources.

The scarcity of public music datasets with directly available audio at scale is intrinsically linked to copyright restrictions, which limit distribution. Although certain legal frameworks, such as the European Union's \cite{EU2019directive}, provide exceptions for research use of copyrighted data, the use of non-openly licensed data for certain tasks, notably generative modeling, still raises significant ethical concerns.

\vspace{0.15cm}
\noindent \textbf{Tagging datasets.}
Early benchmarks such as GTZAN~\cite{tzanetakis2002musical} and MagnaTagATune~\cite{law2009evaluation} laid the foundation for music tagging, but are limited by their small scale, short clip durations, and label inaccuracies~\cite{sturm2013gtzan}. Later datasets, including the Million Song Dataset (MSD)~\cite{bertin2011million}, and AudioSet~\cite{gemmeke2017audio} (for general audio), expanded coverage and diversity, but do not provide raw audio directly. Instead, they rely on links to external sources for audio retrieval, leading to reproducibility challenges due to broken links and content volatility. Other large-scale recent initiatives include metadata-only datasets like Music4all~\cite{santana2020music4all}, DISCO-10M~\cite{lanzendorfer2023disco}, and variants~\cite{ahmed2025sleeping, solak2025bias}. These datasets can reach millions of samples with diverse annotations, but do not curate for openly licensed audio and do not provide links to audio sources.

Datasets such as FMA~\cite{fma_dataset} and MTG-Jamendo~\cite{bogdanov2019mtg} are attempts to address the ethical and legal barriers of copyrights. They provide metadata and full-length music tracks distributed under CC licenses from the Free Music Archive\footnote{\url{https://freemusicarchive.org/}} and Jamendo.\footnote{\url{https://www.jamendo.com}} However, their scale remains limited relative to the demands of recent SOTA models for music understanding and generation~\cite{copet2023simple, tal2024joint}. Larger CC-licensed datasets, like the AcousticBrainz dataset~\cite{bogdanov2019acousticbrainz} and the CC-subset of Freesound\footnote{\url{https://freesound.org/}} used to train Stable Audio Open~\cite{evans2025stable} require external sources for audio or suffer from content volatility.

\vspace{0.15cm}
\noindent \textbf{Captioning datasets.}
The rise of audio-language multi-modal learning increased the demand for datasets with natural language descriptions of audio. To meet this need, initially several human-annotated music-caption datasets such as MusicCaps, SongDescriber, and YT8M~\cite{agostinelli2023musiclm, manco2023song, mckee2023language} were proposed. Although they provide high-quality captions, these datasets are small scale and primarily serve for evaluation purposes. 

Early automatically annotated caption datasets relied on converting tagging datasets into captioning datasets using heuristics~\cite{schneider2024mousai,evans2025stable} or LLMs~\cite{doh2023lp}. More recent initiatives use LALMs for automated captioning, a strategy that facilitates scaling. While LALMs can be used directly to obtain music descriptions, a two-stage automated captioning pipeline is becoming increasingly popular in the literature~\cite{wu2024futga, vyas2025pushing, roy2025jamendo, lanzendorfer2025coarse}. This strategy typically consists of combining a base caption created by a LALM with tags obtained from a source dataset using an aggregator LLM~\cite{wu2024futga, lanzendorfer2025coarse, vyas2025pushing}. Informed by the tags, the LLM is capable of correcting fine-grained base captions generated by the LALM, resulting in better quality descriptions. Similar approaches include~\cite{salganik2026musicsem}, which relies on real-life Reddit musical discussions as complementary data, and~\cite{melechovsky2024mustango, kumar2025sila}, which add descriptors extracted from computed acoustic or musical attributes to LALM captions.

Among recent initiatives, JamendoMaxCaps (JMC)~\cite{roy2025jamendo} stands out as a significant milestone, offering an order of magnitude more CC-licensed full length audio than previous datasets. JMC specifically addresses the challenge of missing or sparse metadata through a sophisticated two-stage automatic captioning and imputation pipeline. First, base captions for $30$-s music segments in the dataset are generated using Qwen2-Audio~\cite{chu2024qwen2audio}, which is prompted with the raw audio segment and any available partial metadata available in Jamendo for the file from which the segment was extracted. Following this, a retrieval system is designed to impute missing tags from similar songs identified using a combination of MERT~\cite{li2024mert} features, partial text metadata, and LLM post-processing. In JMC, imputed metadata is not used to further refine the base captions. 

Our work builds upon this methodological foundation, but scales dataset size considerably by targeting a larger pool of uncurated content scraped from IA. In addition to this, we incorporate tag imputation using SOTA audio classifiers so as to enhance base captions in a more metadata-scarce setting. Despite filtering to ensure quality and license awareness of our data samples, our resulting dataset exceeds the scale of JMC without losing caption quality, providing the community with a more expansive resource for music understanding and generative modeling research.
\section{The Internet Archive Music Dataset}\label{sec:iamd}
\begin{figure*}[t]
\centering
\small
\begin{tikzpicture}[
    node distance=7mm and 10mm,
    box/.style={draw, rounded corners, align=center, font=\scriptsize, text width=2.8cm},
    io/.style={align=center, font=\scriptsize}, 
    arrow/.style={-Latex, thick}
]

\node[io] (ia) {Internet Archive};
\node[box, right=of ia] (f1) {Media type + CC filter};
\node[box, right=of f1] (llm) {LLM filtering};
\node[box, right=of llm] (dl) {Download\\+ duration filter};

\node[box, below=of ia] (clean) {Deduplication};
\node[box, right=of clean] (mb) {MusicBrainz\\ISWC check};
\node[box, right=of mb] (seg) {30s segmentation\\+ silence removal};
\node[io, right=of seg] (out) {IAMD audio segments};
\draw[arrow] (ia) -- (f1);
\draw[arrow] (f1) -- (llm);
\draw[arrow] (llm) -- (dl);

\draw[arrow] (dl) -- ++(0,-0.5) -| (clean);

\draw[arrow] (clean) -- (mb);
\draw[arrow] (mb) -- (seg);
\draw[arrow] (seg) -- (out);

\end{tikzpicture}
\caption{Overview of the procedure for obtaining IAMD audio segments. The number of files kept after each filtering stage is reported in the companion webpage.}
\label{fig:scraping-pipeline}
\end{figure*}
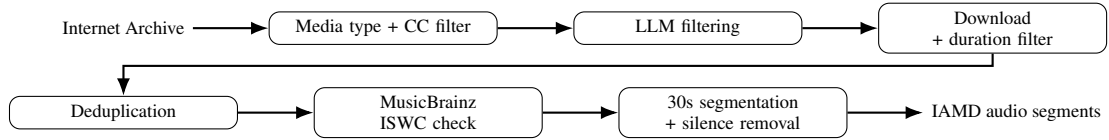

\subsection{The Internet Archive}\label{sec:internet-archive}
The IA is a free-access digital library dedicated to storing web pages and other content uploaded to the Internet since 1996. Today, it houses over a trillion web pages, millions of books, images and videos, and importantly for the MIR community, around $13$ million audio recordings~\cite{archive2014about}.

The IA is crowd-sourced, and anyone with an account can upload content to it. Uploads to the IA are organized in \textit{items} which are further subdivided into \textit{files}. Items are simply collections of files; for example, for a given music album, it might be album covers, tracks in different formats, etc. Uploads have associated metadata, such as author fields and license fields. In the context of audio recordings, item metadata may include useful data such as genre, year of release, and natural language descriptions with mood or instrumentation information.

It is possible to query the IA using a Python API.\footnote{\url{https://archive.org/developers/internetarchive/index.html}} Importantly, this allows one to automatically search the IA for items containing audio recordings declared by users as CC-licensed. Below, we describe our scrapping, cleaning, and segmenting strategy for building IAMD, summarized in Figure~\ref{fig:scraping-pipeline}.

\subsection{Scraping, Cleaning, and Segmenting}\label{sec:scrapping}

\noindent\textbf{Media type + CC filter.} In addition to the fields mentioned above, all IA uploads have an important metadata field named \textit{media type}. Media types determine if the upload is an audio recording, a movie, an image, or another type of data. In the initial stage of our data collection pipeline, we query the IA API for all items having media type \textit{audio}. We retain items satisfying at least one of the following: (i) having a \textit{license url} metadata field pointing to a Creative Commons license of any kind; or (ii) having a --- capitalized or not, with or without spaces --- occurrence of the keywords ``creative commons'' in its description. Then, items with license URLs pointing to CC-BY-NC-ND licenses are excluded from the collection, as no-derivatives licenses do not allow modification of the original data, which could include deep learning model training in the case of generative models. We note however that an important limitation of this filter is that licenses are self-declared by users and may contain errors.

\vspace{0.15cm}
\noindent\textbf{LLM filtering.} In a second filtering stage, we aggregate all available metadata for each item, including free-form fields such as descriptions and reviews, and feed it to a local LLM. The model is prompted to assign a score from 1 (least likely) to 5 (most likely) indicating whether the item contains musical audio. Items with scores greater than 4 are retained. This stage is designed to filter out non-musical content (e.g., podcasts, radio broadcasts) based on semantic cues in the metadata. We use the LLaMA-3.1-8B-Instruct model for this task, with the prompt provided on our companion page.

\vspace{0.15cm}
\noindent\textbf{Download + duration filter.} Items remaining in the collection are then downloaded. For each item, we prioritize the FLAC format; if unavailable, we fall back to AAC, Vorbis, or MP3, in this order of preference. When downloading lossy encoded files, we always use the highest bitrate available. Statistics on file formats for IAMD can be found in our companion page. We then check the duration of the downloaded files. Radio programs, TV shows, and podcasts tend to have longer durations than most music files. It is common, for example, to find music with a duration of around $3$ minutes, whereas it is rare to find a podcast of the same duration. Based on this we keep only files having a duration of less than $10$ minutes for practical reasons, at the cost of potentially underrepresenting long-form musical styles such as jazz and classical.

\vspace{0.15cm}
\noindent\textbf{Deduplication.} After this, a fourth filtering stage is applied to remove duplicate files. The list of remaining files is sorted by filename and increasing duration, and consecutive entries with duration differences smaller than two seconds are treated as duplicates. In such cases, the bottom entry is removed. This heuristic is motivated by the observation that duplicate files often share nearly identical filenames and durations, differing only due to encoding variations.

\vspace{0.15cm}
\noindent\textbf{MusicBrainz ISWC check.} To enable a more conservative assessment of user-declared licenses, we attempt to cross-reference the remaining files with MusicBrainz. For each file, artist names are extracted from the IA metadata and matched to MusicBrainz entries using fuzzy string matching. When a match is found, the filename is similarly matched to identify potential correspondences. Files associated with matched entries containing registered ISWCs\footnote{\url{https://www.iswc.org/}} are flagged as very likely copyrighted. In addition, files whose matched artists have works with registered ISWCs are marked as potentially non-CC. Files without reliable matches retain their original IA licensing information.

\vspace{0.15cm}
\noindent\textbf{Segmentation + silence removal.} The remaining files are segmented into non-overlapping $30$-s segments. This enables multiple segment-level annotations to be extracted from each file while providing finer temporal granularity, as different parts of a piece may vary in instrumentation, mood, or even genre. If a file cannot be divided into $30$-s segments, the remainder at the end of the file is discarded while the segment at the beginning is kept. In a final filtering stage, all segments having an RMS energy below $-50$\,dB are considered silent and discarded. This threshold was chosen based on the energy distribution of the segments, available on the companion page for this dataset. With this criterion, approximately $2\%$ of the final segments were removed. The remaining segments constitute the audio entries of the IAMD. All metadata of their parent files and items is kept for future use in the automatic captioning pipeline.

\subsection{Base Captioning}
\begin{figure*}[t]
    \centering
    \includegraphics[width=\textwidth]{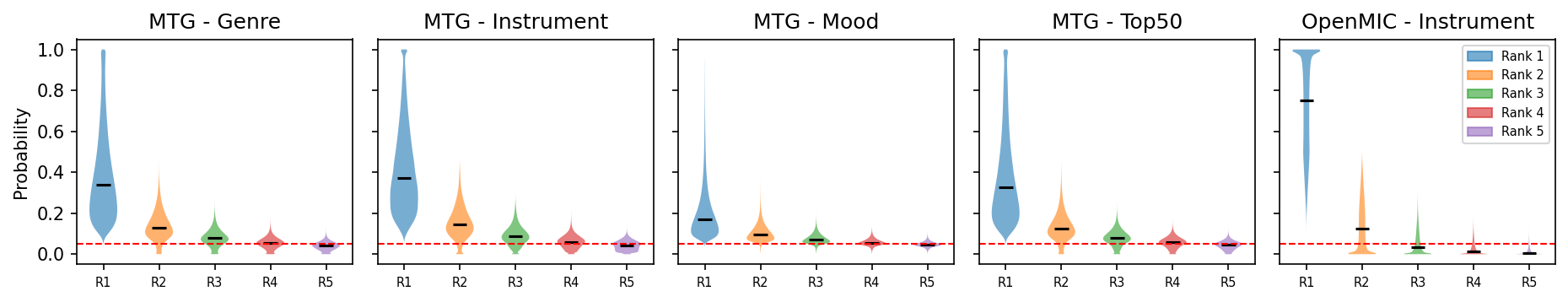}
    \caption{Distribution of tag probabilities per-rank for MATPAC++ classifiers. Red dotted line at $5\%$.}
    \vspace{-0.3cm}
    \label{fig:tag-probabilities}
\end{figure*}

After extracting music segments from the original audio files, we generate a base caption for each segment using an audio-language model. These base captions are later enhanced with additional information obtained from ground truth and imputed metadata, and aggregated into a final caption using an LLM, as in previous works~\cite{wu2024futga, vyas2025pushing, lanzendorfer2025coarse}

To obtain the base caption for each segment, we use TinyMU~\cite{li2026tiny}, a lightweight music understanding model. TinyMU achieves music understanding performance comparable to significantly larger LALMs, such as AudioFlamingo3~\cite{goel2025audioflamingo} and Qwen2-Audio~\cite{chu2024qwen2audio}. It notably achieves $99\%$ of Qwen2-Audio's BERTScore on the test set of MusicCaps~\cite{agostinelli2023musiclm}, while having only 229M parameters (vs.~$8.4$B for Qwen2-Audio).

The choice of TinyMU over more recent LALMs~\cite{xu2025qwen3omni, goel2025audioflamingo, ghosh2026music} balances computational efficiency with sufficient base caption quality. While larger models could improve initial captions, they would incur substantially higher computational cost. In our pipeline, this stage is not intended to produce final descriptions, but rather weak first-stage annotations that are later refined by an LLM. Given TinyMU’s competitive performance, the large number of segments to process, and the subsequent refinement step, we adopt it as an efficient initial annotator.

The prompt given to TinyMU in order to generate the base captions can be found in the dataset's companion page. In addition to this, the base captions for all segments in the dataset are made available along with the final captions, enabling direct comparison between intermediate and refined annotations.

\subsection{Metadata Extraction}
\begin{figure*}
    \centering
    \includegraphics[width=0.95\textwidth]{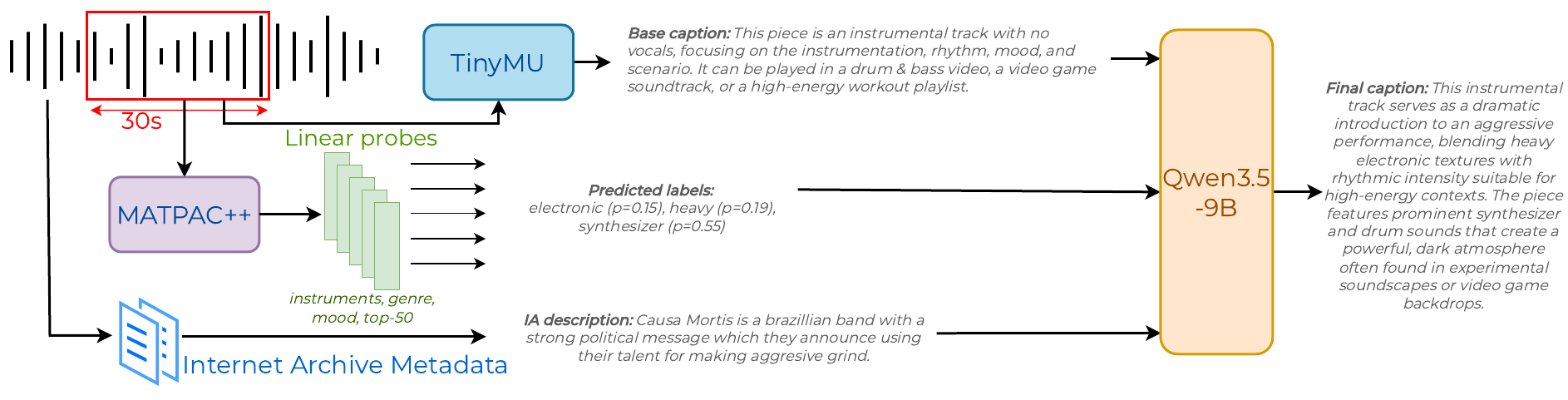}
    \vspace{-0.3cm}
    \caption{Complete caption generation pipeline with an example of generated caption.}
    \label{fig:iamd_diagram}
\end{figure*}

To build IAMD, base captions are enhanced with two additional sources of metadata: textual information from IA uploads, and audio-derived tags obtained by audio classifiers based on MATPAC++~\cite{quelennec2025matpacenhanced}. These provide complementary signals based on user-provided audio descriptors and learned audio features, respectively.

\vspace{0.15cm}
\noindent \textbf{Textual descriptors.}
IA uploads have metadata that can be useful for captioning. This is information provided by users at upload time, and as such, it must be noted that it is not entirely reliable. Despite this limitation, IA metadata is a valuable source of human annotations that can be used to improve upon the first stage of the captioning pipeline. 

File-level metadata is mostly checksums or information unrelated to musical attributes. Item-level metadata, on the other hand, may contain relevant musical information in the form of structured tags (rare) or free-form descriptions (common). Natural language user descriptions are kept without modification to complement the base captions. We refer the reader to the companion page to the dataset for a detailed description of available IA metadata, including examples of free-form upload descriptions. 

\vspace{0.15cm}
\noindent \textbf{Audio classification tags.}
Segment-level annotations are essential to our dataset, as they provide information at finer temporal resolution for caption generation. To obtain segment-level information, we use MATPAC++ audio classifiers. MATPAC++ is a self-supervised audio representation model that supports effective downstream tagging via linear probing, achieving state-of-the-art or competitive performance compared to other SSL and specialized tagging models~\cite{li2024mert, quelennec2025matpacenhanced}.

We used five classifiers, trained on the OpenMIC dataset instrument tags, and on the MTG-Jamendo instrument, genre, mood, and top-50 tags. For each segment, we obtained the five most likely tags for each of the classifiers, and then plotted the distribution of tag probabilities across rank positions, i.e., the distribution of tag probabilities for the most likely tag, second most likely tag, and so on. This was done for all classifiers and is displayed in Figure~\ref{fig:tag-probabilities}.

Looking at these distributions, it is possible to see that, for all classifiers, from the third most likely tag onward, probabilities are concentrated below $10\%$, with most of the values around $5\%$. Based on this, we applied a conservative threshold and further eliminated any output tags having probabilities below $5\%$.

\subsection{Final Caption Generation}\label{sec:final-captioning}
Our captioning pipeline, summarized in Figure~\ref{fig:iamd_diagram}, enhances the segment base captions obtained from TinyMU with the textual information extracted from the corresponding IA uploads and the tags obtained by MATPAC++ classifiers. At its last stage, a structured prompt is used to ask a LLM to combine base captions and additional information in order to obtain the final captions.

In this two-stage pipeline, the LLM acts as an aggregator; it takes into account the additional information to correct errors in the base captions and introduce new details. We use Qwen3.5-9B\footnote{\url{https://hf.co/Qwen/Qwen3.5-9B}} as our aggregator LLM. It offers strong reasoning performance while remaining efficient enough for local execution, making it a practical choice for large-scale processing where larger models would be computationally prohibitive.

The prompting strategy to obtain the final captions asks Qwen3.5-9B to retain the core meaning of the base caption while refining genre, mood, and instrumentation information using the MATPAC++ linear probes and the textual metadata. Linear probe tags are provided along with their corresponding probabilities, in order to have reliability information encoded along with the tags. The LLM is asked to ignore ambiguous signals, retaining general descriptions in case of conflict between information sources. Finally, it is instructed to generate captions that are at most two or three sentences long to keep descriptions concise. The full prompt given to Qwen3.5-9B is available on the dataset's companion page, which also reports runtime statistics and hardware usage for each stage of the dataset construction and caption generation pipeline.

\begin{table*}[!t]
    \small
    \centering
    \setlength{\tabcolsep}{4pt}
    \renewcommand{\arraystretch}{0.95}
    \begin{tabular}{lccccc}
        \toprule
        & \multicolumn{2}{c}{\textbf{CAF (\%)}} & \multicolumn{2}{c}{\textbf{S-CLAP (\%)}} & \textbf{FLEUR (\%)} \\
        \cmidrule(lr){2-3} \cmidrule(lr){4-5}
        & LAION & M2D & LAION & M2D & \\
        \midrule
        \multicolumn{6}{l}{\textbf{30s segments}} \\
        \midrule
        JMC  & $52.29 \pm 0.66$ & $35.43 \pm 0.36$ & $47.91 \pm 0.60$ & $26.84 \pm 0.19$ & $\mathbf{69.80 \pm 1.44}$ \\
        IAMD & $\mathbf{55.61 \pm 0.77}$ & $\mathbf{35.50 \pm 0.44}$ & $\mathbf{53.15 \pm 0.69}$ & $\mathbf{28.01 \pm 0.20}$ & $65.43 \pm 1.77$ \\
        \midrule
        \multicolumn{6}{l}{\textbf{10s segments}} \\
        \midrule
        MusicCaps & $\mathbf{56.06 \pm 0.53}$ & $\mathbf{39.36 \pm 0.27}$ & $\mathbf{49.07 \pm 0.59}$ & $\mathbf{28.18 \pm 0.21}$ & $\mathbf{84.06 \pm 0.95}$ \\
        $\mathrm{IAMD}_{mid}$ & 50.32 $\pm$ 0.84 & 34.65 $\pm$ 0.47 & 46.41 $\pm$ 0.75 & 26.82 $\pm$ 0.21 & 65.96 $\pm$ 1.86 \\
        $\mathrm{IAMD}_{rnd}$ & 50.26 $\pm$ 0.84 & 34.61 $\pm$ 0.47 & 46.34 $\pm$ 0.75 & 26.78 $\pm$ 0.21 & 65.92 $\pm$ 1.87 \\
        \bottomrule
    \end{tabular}
    \caption{Reference-free caption quality metrics for IAMD compared to JMC (30-s segments) and MusicCaps (10-s segments). CAF and S-CLAP scores are reported using two backbones. Values are sample means with 95\% CIs, computed over 1500 randomly sampled segments. Best in bold.}
    \vspace{-0.3cm}
    \label{tab:iamd-caf}
\end{table*}

\section{Evaluation}\label{sec:evaluation}
\subsection{Objective Evaluation}
To assess the quality of the generated captions, we conducted an objective evaluation using the reference-free CAF-Score~\cite{lee2026caf} as our main caption quality metric.

The CAF-Score, or CLAP-aligned FLEUR score, is a combination of a CLAP similarity score and a FLEUR score~\cite{lee2024fleur}. CLAP-based metrics measure whether the caption describes semantic attributes present in the audio (relevant for downstream text-to-audio tasks), but do not assess linguistic fluency or readability. FLEUR scores, on the other hand, use the output logits of a LALM prompted to evaluate a caption. As such, they evaluate fluency and coherence (relevant for downstream language modeling), but are vulnerable to model hallucinations and biases.

CAF-Scores are calculated as a linear interpolation of S-CLAP, the maximum cosine similarity between the CLAP embedding of the caption and CLAP embeddings computed over sliding windows of the audio, and FLEUR. We use the reference implementation,\footnote{\url{https://github.com/inseong00/CAF-Score}} in which S-CLAP is truncated to the $[0,1]$ range. This results in CAF-Scores also lying in $[0,1]$, as FLEUR lies in the same range. In~\cite{lee2026caf} the authors show evidence that CAF-Scores are more correlated to human preferences than individual S-CLAP and FLEUR scores, and that in some cases CAF-Scores can be almost as similarly correlated to human preferences as reference-based metrics.

Following~\cite{lee2026caf}, Qwen3-Omni-30B was used to obtain FLEUR scores. To obtain S-CLAP scores, we used both LAION-CLAP,\footnote{\url{https://hf.co/laion/clap-htsat-unfused}} and M2D-CLAP~\cite{niizumi2024m2dclap}. These two models obtain the best correlation with human preferences in~\cite{lee2026caf}. Also following~\cite{lee2026caf}, we combine S-CLAP and FLEUR with weights of $0.8$ and $0.2$, respectively.

Our objective evaluation consists of two comparisons. In the first, we sample $1,500$ segments from both IAMD and JMC and compute mean CAF-Scores using the two CLAP backbones. In the second, we reuse the same $1500$-sample subset of IAMD and compare it against a randomly sampled subset of $1,500$ MusicCaps segments. To ensure comparability, we control for segment duration, as MusicCaps samples are $10$-s long and S-CLAP may favor longer segments due to its sliding-window maximum operation. For the second comparison we therefore crop IAMD segments to $10$-s with two setups: random crop ($\mathrm{IAMD}_{rnd}$) and centered crop ($\mathrm{IAMD}_{mid}$). Results are reported as sample means with $95\%$ confidence intervals (CIs) in Table~\ref{tab:iamd-caf}. We additionally report mean S-CLAP and FLEUR scores with corresponding $95\%$ CIs.

We observe that IAMD and JMC obtain similar results overall. IAMD is slightly better in CAF-Score and S-CLAP, whereas JMC obtains higher FLEUR scores. This suggests that IAMD scales while maintaining competitive caption quality. In the comparison with MusicCaps, IAMD lags behind in all metrics --- which is expected given that MusicCaps is human-labeled --- but remains competitive in CAF-Scores and S-CLAP. Differences in FLEUR scores in both cases suggest that linguistic aspects might be a point for improvement of IAMD captions.

\subsection{Subjective Evaluation}

As a complement to the objective evaluation of IAMD captions, we also conducted a subjective test evaluating caption \textit{correctness} and \textit{completeness}. Volunteers were asked to evaluate correctness as the absence of wrong or inaccurate descriptions, i.e., whether the caption includes information that is not supported by the audio. Conversely, they were asked to evaluate completeness as whether or not the caption omits relevant information that is present in the audio. For this evaluation, we defined relevant information as descriptors of musical genre, mood, and instrumentation.

Volunteers rated both aspects on a continuous scale from one to five with discrete anchors. The anchors were defined as: incorrect/incomplete (1), mostly incorrect/incomplete (2), somewhat correct/complete (3), mostly correct/complete (4), and correct/complete (5). The definitions of correctness and completeness mentioned above were displayed alongside each question. 

The test was conducted online,\footnote{\url{https://ia-db-v0-audio-subjective-test.onrender.com/}} and participants were instructed to complete it in a quiet environment using headphones. The evaluation consisted of five warm-up questions (identical for all participants) using MusicCaps captions, to familiarize participants with the task using high-quality reference captions, followed by ten questions evaluating IAMD captions. On average, volunteers took $17$ minutes to complete the test.
\begin{table}[t]
    \small
    \centering
    \setlength{\tabcolsep}{5pt}
    \begin{tabular}{cc}
        \toprule
        \textbf{Correctness} & \textbf{Completeness} \\
        \midrule
        4.24 $\pm$ 0.13 & 4.35 $\pm$ 0.11 \\
        \bottomrule
    \end{tabular}
    \caption{Subjective evaluation of IAMD captions. Scores for correctness and completeness are reported on a 1--5 scale as per-sample means with $95\%$ CIs.}
    \vspace{-0.3cm}
    \label{tab:subjective-iamd}
\end{table}

A random selection of $100$ samples from IAMD constituted the pool of signals for the test. A total of $20$ volunteers took the test. A circular shift strategy was used to ensure that each signal in the pool was evaluated exactly twice, that no two tests were identical, and that no signal appeared more than once in the same test. Signals were presented to volunteers in a random order to avoid bias.

Results were first aggregated per sample by averaging ratings across participants, and then summarized over the full set of test samples. The mean scores for correctness and completeness are reported in Table~\ref{tab:subjective-iamd}, together with $95\%$ CIs computed over samples. The results show that IAMD captions achieve consistently high scores on both evaluation criteria, indicating good semantic quality of the captions.
\section{Conclusion}\label{sec:conclusion}
With over 34,000 hours of audio, the IAMD dataset is, to the best of our knowledge, the largest music-caption dataset to date. According to our objective evaluation, its captions show quality comparable to JMC, the largest automatically annotated music caption dataset until now, suggesting scaling without caption quality loss. Furthermore, objective results for IAMD do not lag far behind MusicCaps, which is human-annotated. 

A subset of IAMD can be linked to the MusicBrainz database, providing a basis for more conservative verification of licensing information. In the future, samples having hits in MusicBrainz could be used to enrich the dataset with high-quality structured tags.

We release the code for data collection, processing, and segmentation of IA content, as well as the cross-referencing pipeline and the two-stage captioning framework, to support reproducibility and future extensions of this work. We hope that IAMD will serve as a large-scale benchmark for music understanding and generative modeling tasks, and contribute to more reproducible research in music information retrieval.
\section{Ethics Statement}
The IAMD dataset is constructed from publicly available audio content and is intended strictly for research purposes. It aims at improving access to large-scale music–caption data for the academic community, addressing reproducibility challenges associated with datasets that rely on external links or restricted access. By releasing both the audio data and the associated processing and annotation pipelines, we promote transparency and enable future work to build upon and improve the dataset. We note that while the dataset relies on content declared under Creative Commons licenses, metadata inaccuracies may remain, and users are responsible for ensuring compliance with applicable copyright laws across different legal frameworks.

We acknowledge that large-scale music datasets may have broader societal and cultural implications, particularly in the context of generative modeling. Such models can enable new forms of artistic expression and research, but may also raise concerns regarding authorship, fair compensation, and the potential homogenization of musical styles. We encourage the responsible use of IAMD with awareness of these considerations and its potential downstream impact on creative and cultural domains.

\bibliography{references}

\end{document}